\documentclass[journal,twoside]{IEEEtran}
\usepackage{cite}
\ifCLASSINFOpdf
\else
   \usepackage[dvips]{graphicx}
\fi

\usepackage{amsfonts}
\usepackage{lpic}
\usepackage{psfrag}
\usepackage{epsf} 
\usepackage{dsfont}
\usepackage{textcomp}
\usepackage{stmaryrd}
\usepackage{amssymb}
\usepackage{mathrsfs}
\usepackage{textcomp}
\usepackage{graphicx}
\usepackage[belowskip=-11pt,aboveskip=10pt,footnotesize]{caption}
\usepackage{caption}
\usepackage{subcaption}
\usepackage{amsbsy}
\usepackage{lettrine}
\newcommand{\tr}[1]{\mathrm{#1}}

\newcommand{\diag}{\mathop{\mathrm{diag}}}

\usepackage[cmex10]{amsmath}
\usepackage{accents}
\usepackage{array}
\usepackage{mdwmath}
\usepackage{mdwtab}

\begin{document}
%
\title{Estimation of Phase Noise \\in Oscillators with Colored Noise Sources}

\author{\hspace{0in}\IEEEauthorblockN{M. Reza Khanzadi, \emph{Student Member, IEEE}, Rajet Krishnan, \emph{Student Member, IEEE}, and Thomas Eriksson}
\thanks{The authors are with the Department of
Signals and Systems, Chalmers University of Technology, Gothenburg, Sweden. M. Reza Khanzadi is also with the Department of Microtechnology and Nanoscience, Chalmers University of Technology. Email: \{khanzadi, rajet, thomase\}@chalmers.se.}
\vspace{-0.6cm}}

\markboth{IEEE COMMUNICATIONS LETTERS}{KHANZADI \MakeLowercase{\textit{et al.}}: Estimation of Phase Noise in Oscillators with Colored Noise Sources}
\maketitle
\begin{abstract}
In this letter we study the design of algorithms for estimation of phase noise (PN) with colored noise sources. A soft-input maximum a posteriori PN estimator and a modified soft-input extended Kalman smoother are proposed. The performance of the proposed algorithms are compared against those studied in the literature, in terms of mean square error of PN estimation, and symbol error rate of the considered communication system. The comparisons show that considerable performance gains can be achieved by designing estimators that employ correct knowledge of the PN statistics.
\vspace{-0.4cm}
\end{abstract}
\IEEEpeerreviewmaketitle
\section{Introduction}
\label{Sec_Introduction}
\lettrine[lines=2,nindent=0pt]{O}SCILLATOR PHASE NOISE (PN) results in challenging synchronization issues which degrade the performance of communication systems \cite{Colavolpe2005, Mehrpouyan2012}. Demands for high data rates motivate the use of high-order modulation schemes in such systems. Nevertheless, PN severely limits the performance of systems that employ dense constellations.

The problem of PN estimation has been widely studied during the last decades (see \cite{Amblard2003151, Colavolpe2005, Bhatti2009} and the references therein). In \cite{Bhatti2009}, a feedforward PN estimation-symbol detection algorithm is presented, while iterative methods for joint phase estimation and symbol detection are studied in \cite{Colavolpe2005}.

In prior studies, PN is modeled as a discrete random walk with uncorrelated (white) Gaussian increments between each time instant (i.e., the discrete Wiener process). This model results from using oscillators with white noise sources \cite{Demir2006}. However, numerous studies show that real oscillators also contain colored noise sources, and PN is accurately modeled as a random walk with correlated (colored) Gaussian increments \cite{Khanzadi2012_1, Khanzadi2013_1, Demir2006}.

In this letter, we propose techniques to estimate PN from real oscillators with white and colored noise sources, in a single antenna-single carrier communication system. We first derive a general soft-input maximum a posteriori (MAP) PN estimator that is optimal in terms of the mean square error (MSE). Then, a modified soft-input extended Kalman smoother is proposed that can be used for estimation of PN with colored increments. The proposed Kalman smoother is observed to perform close to the MAP estimator in several interesting scenarios, with a significantly reduced complexity. Further, we compare the proposed methods with state of the art techniques. The proposed estimators jointly estimate the PN samples of a block of received signals, which improves the estimation performance compared to sequential PN estimation algorithms previously studied (e.g., \cite{Amblard2003151}). Our estimators can be used in feedforward or iterative designs for the estimation of PN with white and colored increments.

\section{System Model}
\label{Sec_System_Model}
Consider the transmission of a block of $K$ data symbols over an additive white Gaussian noise (AWGN) channel, affected by random PN. The channel coefficient from the transmitter to receiver antenna is assumed to be constant over the transmitted block, and it is estimated and compensated by employing a known training sequence that is transmitted prior to the data symbols \cite{Mehrpouyan2012}. In the case of perfect timing synchronization, the received signal after sampling the output of matched filter can be modeled as in \cite{Colavolpe2005}
\vspace{-0.2cm}
\begin{align}
\label{System_Model1}
y_k&=s_ke^{j\theta_k}+w_k,~k\in\{1,\dots,K\},
\end{align}
where $\theta_k$ represents the PN affecting the $k$th received signal due to noisy transmitter and receiver local oscillators, and $w_k$ is a realization of the independent and identically distributed (i.i.d.) zero-mean complex circularly symmetric AWGN with variance $\sigma^2$. In this model, $\mathbf{y}=\{y_k\}_{k=1}^K$ is the sequence of received signals and $\mathbf{s}=\{s_k\}_{k=1}^K$ is the transmitted symbol sequence divided in two sets of symbols, $K_p$ known pilot symbols, and $K-K_p$ unknown data symbols. We model the pilot and data symbols in general as 
\vspace{-0.2cm}
\begin{align}
\label{softSymbolDefine}
s_k=\hat{s}_k+\epsilon_k,
\end{align}
where $\hat{s}_k$ is the soft detected symbol and $\epsilon_k$ models the uncertainty of $s_k$ as an i.i.d zero-mean circularly symmetric AWGN with variance $\sigma^2_{\epsilon_k}$. Such a modeling choice is commonly used in the literature \cite{song2004soft}. For the pilot symbols, $\sigma^2_{\epsilon_k}=0$ since they are known. Using (\ref{System_Model1}) and (\ref{softSymbolDefine}), the received signal can be rewritten as 
\vspace{-0.3cm}
\begin{align}
\label{System_Model1_Modified}
y_k&=\hat{s}_ke^{j\theta_k}+\underbrace{\epsilon_ke^{j\theta_k}+w_k}_{\triangleq\tilde{w}_k},~k\in\{1,\dots,K\},
\end{align}
where $\tilde{w}_k$ is the new observation noise. As $\epsilon_k$ is modeled circularly symmetric, $\tilde{w}_k \sim \mathcal{CN}(0,\sigma_k^2\triangleq\sigma^2+\sigma^2_{\epsilon_k})$.{\let\thefootnote\relax\footnote{\emph{Notations:} Italic letters $(x)$ are scalar variables, boldface letters $(\mathbf{ x})$ are vectors, uppercase boldface letters $(\mathbf{X})$ are matrices, $([\mathbf{X}]_{a,b})$ denotes the $(a,b)^{th}$ entry of matrix $\mathbf{X}$, $\tr{diag}(\mathbf{X})$ denotes the diagonal elements of matrix $(\mathbf{X})$, $\mathds{E}[\cdot]$ denotes the statistical expectation operation, $\mathcal{CN}(x;\mu,\sigma^2)$ denotes the complex proper Gaussian distribution with variable $x$, mean $\mu$, and variance $\sigma^2$, $\log(\cdot)$ denotes the natural logarithm, $\Re\{\cdot\}$, $\Im\{\cdot\}$, and $\arg\{\cdot\}$ are the real part, imaginary part, and angle of complex-valued numbers, and $(\cdot)^*$ and $(\cdot)^T$ denote the conjugate and transpose, respectively.}}\par

The PN samples are modeled by a random-walk as
\vspace{-0.2cm}
\begin{align}
\label{Random_Walk}
\theta_k=\theta_{k-1}+\zeta_{k-1},
\end{align}
where the phase increment process $\zeta_{k}$ is a zero-mean Gaussian random process. Recent studies of the PN in oscillators with colored noise sources show that the PN increments can be correlated over time \cite{Demir2006, Khanzadi2012_1,Khanzadi2013_1}. Hence, we consider a general case where the autocorrelation function of $\zeta_{k}$, denoted as $R_\zeta(l)$, is known a priori. Note that the Wiener PN model extensively used in the literature is a special case of the proposed model, with uncorrelated (white) phase increments \cite{Demir2006}.

\section{Phase Noise Estimation}
\label{Sec_Proposed_MAP_Estimator}
In the sequel, we propose two methods for joint estimation of $K$-dimensional PN vector $\boldsymbol{\theta}=\{\theta_k\}_{k=1}^K$, that further would be used for data detection. First, we derive a MAP estimator. Thereafter, we propose an approach for modification of (\ref{Random_Walk}), such that smoothing algorithms (e.g., Kalman smoother) with a lower complexity than MAP can be used for estimation.  
\vspace{-0.13cm}
\subsection{Proposed MAP Estimator}
\label{SSec_Proposed_MAP_Estimator}
Let $f(\boldsymbol {\theta}|\mathbf{y})$ denote the a posteriori distribution of PN vector $\boldsymbol {\theta}$, given the observation vector $\mathbf{y}$. The MAP estimator of $\boldsymbol {\theta}$ is determined as
\begin{align}
\label{MAP_def}
\hspace{-0.3cm}\boldsymbol {\hat{\theta}}&=\underset{\boldsymbol {\theta}}{\arg \max}\hspace{0.05cm} 
\log (f(\boldsymbol {\theta}|\mathbf{y}))\hspace{-0.05cm}=\hspace{-0.05cm}\underset{\boldsymbol {\theta}}{\arg \max}\hspace{0.05cm} \log (f(\mathbf{y}|\boldsymbol {\theta})f(\boldsymbol {\theta})),\hspace{-0.1cm}
\end{align}
where we define $\ell(\boldsymbol{\theta}) \triangleq \log (f(\mathbf{y}|\boldsymbol {\theta})f(\boldsymbol {\theta}))$. To solve this optimization, we first need to find the likelihood, $f(\mathbf{y}| \boldsymbol {\theta})$, and prior distribution of $\boldsymbol {\theta}$, $f(\boldsymbol {\theta})$. As both $w_k$ and $\epsilon_k$ are i.i.d, and $y_k$ only depends on $\theta_k$ according to (\ref{System_Model1_Modified}), the likelihood function can be written as
\begin{align}
f(\mathbf{y}|\boldsymbol {\theta})=\prod_{k=1}^Kf(y_k|\boldsymbol{\theta})=\prod_{k=1}^Kf(y_k|\theta_k),
\vspace{-0.5cm}
\end{align}
where
\vspace{-0.3cm}
\begin{align}
\label{PN_likelihood}
f(y_k|\theta_k)&=\mathcal{CN}(y_k;\hat{s}_ke^{j\theta_k},\sigma_k^2)\nonumber\\
&=\frac{1}{\sigma_k^2\pi}\exp{\left(-\frac{|y_k-\hat{s}_ke^{j\theta_k}|^2}{\sigma_k^2}\right)}.
\end{align}

In order to find the prior distribution $f(\boldsymbol {\theta})$ of the PN vector, we use the random walk model in (\ref{Random_Walk}) that results in a general PN incremental form of
\vspace{-0.4cm}
\begin{align}
\label{innovation_sum}
&\theta_k=\theta_1+\sum_{i=1}^{k-1}\zeta_i,
\end{align}
where $\theta_1$ (PN of the first symbol in the block) is modeled as a zero-mean Gaussian random variable with a high variance\footnote[1]{We consider a flat non-informative prior \cite{book_kay_est} for the initial PN value, modeled by a Gaussian distribution with a high variance that behaves similar to a flat prior over a certain interval.}, denoted as $\sigma^2_{\theta_1}$. Based on (\ref{innovation_sum}), we can show that $\boldsymbol {\theta}$ has a multivariate Gaussian distribution $f(\boldsymbol {\theta})=\mathcal{N}(\boldsymbol {\theta};\mathbf{0},\mathbf{C})$ where elements of the covariance matrix $\mathbf{C}$ can be computed as
\begin{align}
\label{C_def}
[\mathbf{C}]_{m,m'}&=\mathbb{E}\Big[(\theta_m-\mathbb{E}[\theta_m])(\theta_{m'}-\mathbb{E}[\theta_{m'}])\Big]\\
&=\sigma^2_{\theta_1}+\sum_{l=1}^{m-1}\sum_{l'=1}^{m'-1}R_\zeta(l-l')
,\nonumber \quad m,m'\in\{1,\dots,K\}.
\end{align}
From (\ref{PN_likelihood}) and the multivariate Gaussian prior of $\boldsymbol {\theta}$ we obtain
\begin{align}
\label{log_l}
\ell(\boldsymbol{\theta})
&=\sum_{k=1}^K\frac{2}{\sigma^2_k}\Re\{y_k \hat{s}_k^*e^{-j\theta_k}\}-\frac{1}{2}(\boldsymbol{\theta}^T\mathbf{C}^{-1}\boldsymbol{\theta})+const.
\end{align}
To find the maximizer of (\ref{log_l}), an exhaustive grid-search over all possible values of $\boldsymbol{\theta}$ can be used. However, the complexity of this method increases exponentially with the length of $\boldsymbol {\theta}$. The stationary point of this optimization can analytically be found as the root of the gradient of $\ell(\boldsymbol{\theta})$ with respect to $\boldsymbol {\theta}$,
\begin{align}
\label{d1_log_l}
\mathbf{g}(\boldsymbol{\theta})\triangleq
2\Big[\Big\{\frac{\Im\{y_k \hat{s}_k^*e^{-j\theta_k}\}}{\sigma^2_{k}}\Big\}_{k=1}^K\Big]^T-\mathbf{C}^{-1}\boldsymbol{\theta}=\mathbf{0}.
\end{align}
In order to solve $ \mathbf{g}(\boldsymbol{\theta})= \mathbf{0}$, which is a non-linear system of equations, we use the Newton--Raphson method whose $n$th iteration is given by
\vspace{-0.15cm}
\begin{align}
\label{Newton_Method}
\boldsymbol{\hat{\theta}}^{(n+1)}=\boldsymbol{\hat{\theta}}^{(n)}-\mathbf{H}^{-1}(\boldsymbol{\hat{\theta}}^{(n)})\mathbf{g}(\boldsymbol{\hat{\theta}}^{(n)}),
\end{align}
where $\mathbf{H}(\boldsymbol{{\theta}})$ denotes the Hessian matrix, 
\begin{align}
\label{Hessian_matrix}
\mathbf{H}(\boldsymbol{\theta})\triangleq-\hspace{-0.12cm}\left(\hspace{-0.1cm}2\diag\left(\hspace{-0.12cm}\Big[\Big\{\frac{\Re\{y_k \hat{s}_k^*e^{-j\theta_k}\}}{\sigma^2_k}\Big\}_{k=1}^K\Big]\hspace{-0.12cm}\right)+\mathbf{C}^{-1}\right).
\end{align}
We iterate till an accurate value of the root is reached. 

To show that this algorithm reaches a global maximum, we first prove that $\ell(\boldsymbol{\theta})$ is a concave function in moderate and high signal to noise ratio (SNR) regimes. In (\ref{Hessian_matrix}), $\mathbf{C}^{-1}$ is the inverse of a covariance matrix, thus it is a positive-definite matrix. If the first term of the sum in (\ref{Hessian_matrix}) is also positive-definite (or positive-semidefinite), the Hessian becomes negative-definite, thus implying that $\ell(\boldsymbol{\theta})$ is a concave function. For the first term of (\ref{Hessian_matrix}) to be positive-semidefinite, $\Re\{y_k \hat{s}_k^*e^{-j\theta_k}\}$ must be greater than or equal to zero. Exploiting Eq.~(\ref{System_Model1_Modified}) we get
\vspace{-0.15cm}
\begin{align}
\label{Conc_Proof_1}
\Re\{y_k \hat{s}_k^*e^{-j\theta_k}\}=|{\hat{s}}_k|^2+\Re\{\hat{s}_k^*\check{w}_k\}\geq 0,
\end{align}
where $\check{w}_k\triangleq  e^{-j\theta_k} \tilde{w}_k$ with the same statistics as $\tilde{w}_k$. It is clear that for moderate and high SNR, (\ref{Conc_Proof_1}) is satisfied with a high probability, and consequently $\ell(\boldsymbol{\theta})$ is a concave function. 

In the low-SNR regime, $\ell(\boldsymbol{\theta})$ is not necessarily concave. Therefore, we propose an approach to initiate the iterations with a guess which is fairly close to the optimal point. This ensures that the method does not get trapped in a local maxima, far from the global maximum. Moreover, employing a good initial guess speeds up the convergence of the algorithm at any SNR. In this respect, we first find the maximum likelihood (ML) estimate of the PN samples for the pilot symbols. For any $s_k$ in the pilot set, the ML estimator of the $k$th PN sample can be computed as $\hat{\theta}_k^\tr{ML}=\arg\{y_k s^*_k\}$. Then, we form our initial estimate of the PN vector, $\boldsymbol{\hat{\theta}}^{(0)}$, as the linear interpolation of the ML estimated PN values.

The MAP estimator is an optimal minimum mean square estimator if its MSE attains the Bayesian Cram\'{e}r-Rao bound (BCRB) \cite{book_kay_est}. In the Appendix, we analytically derive the MSE of the MAP, and show that it is approximately equal to the BCRB of the PN estimation. Our simulation results in Sec.~IV also confirms this (Fig.~\ref{fig:MAP_var}). 

Although, the proposed MAP estimator gives an optimal estimate of $\boldsymbol{\theta}$, as we can see in (\ref{Newton_Method}), it involves inversion of $K \times K$ matrices that may raise some complexity issues. In the next section, we propose an approach to modify the PN model and reduce the complexity.
\vspace{-0.15in}
\subsection{Auto Regressive Model of Colored Phase Noise Increments}
\label{SSec_Smoothing}
\begin{figure*}[!tb]
        \centering
         \begin{subfigure}[b]{0.32\textwidth}
                \centering
                \includegraphics[width=\textwidth]{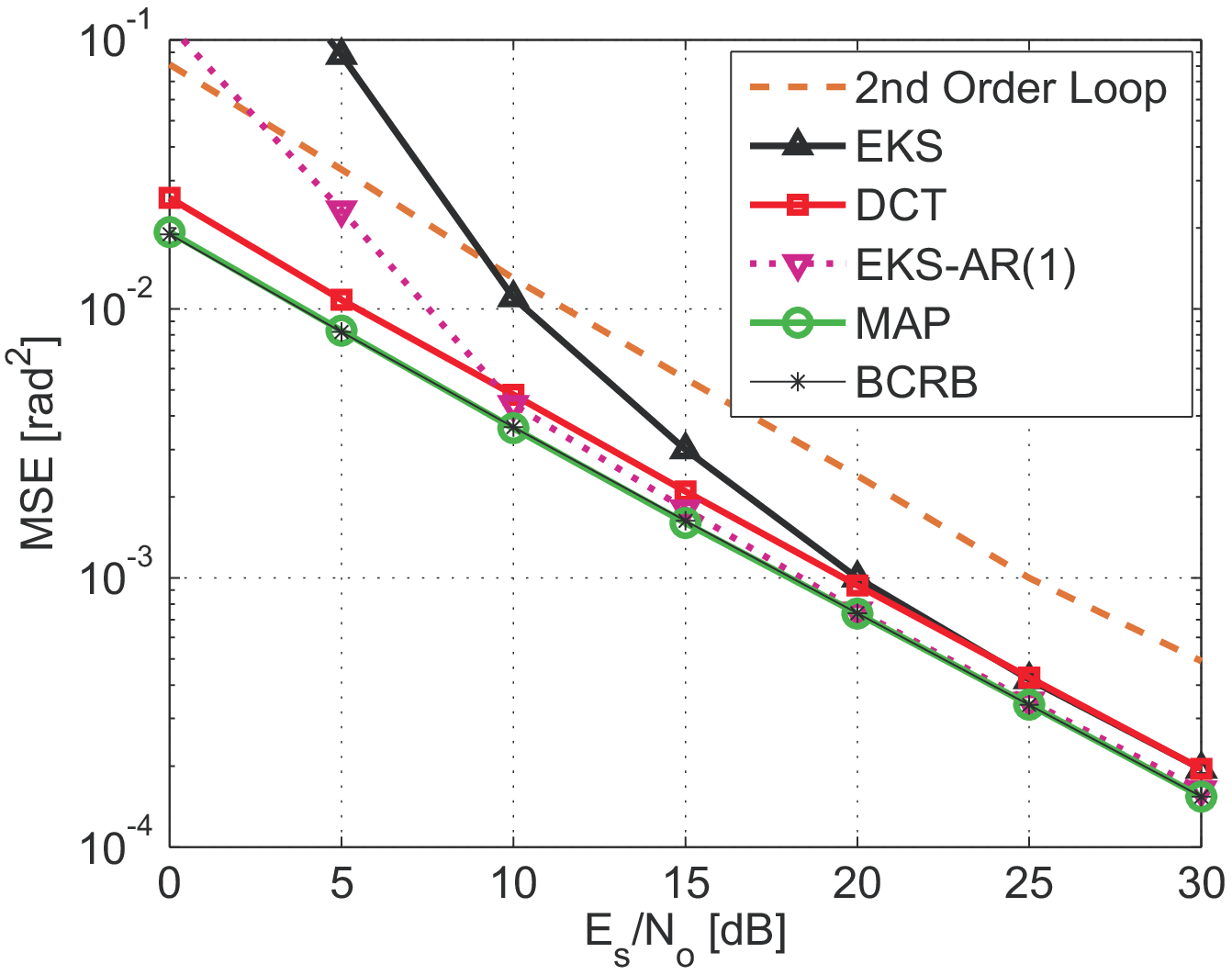}
                \caption{}
                \label{fig:MSE_DA}
        \end{subfigure}
        ~ 
        \begin{subfigure}[b]{0.32\textwidth}
                \centering
                \includegraphics[width=\textwidth]{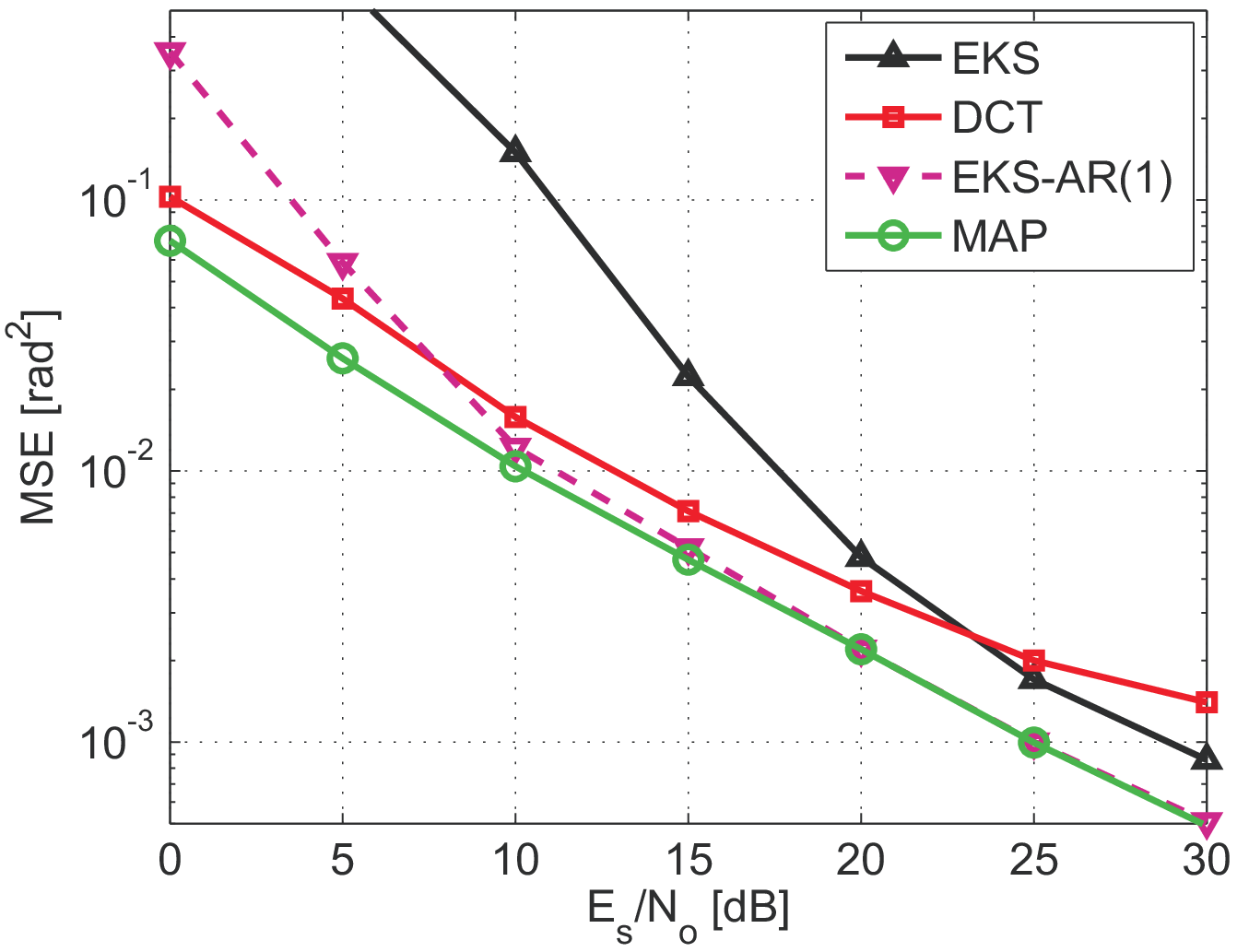}
                \caption{}
                \label{fig:MSE21}
        \end{subfigure}
        ~ 
		 \begin{subfigure}[b]{0.32\textwidth}
                \centering
                \includegraphics[width=\textwidth]{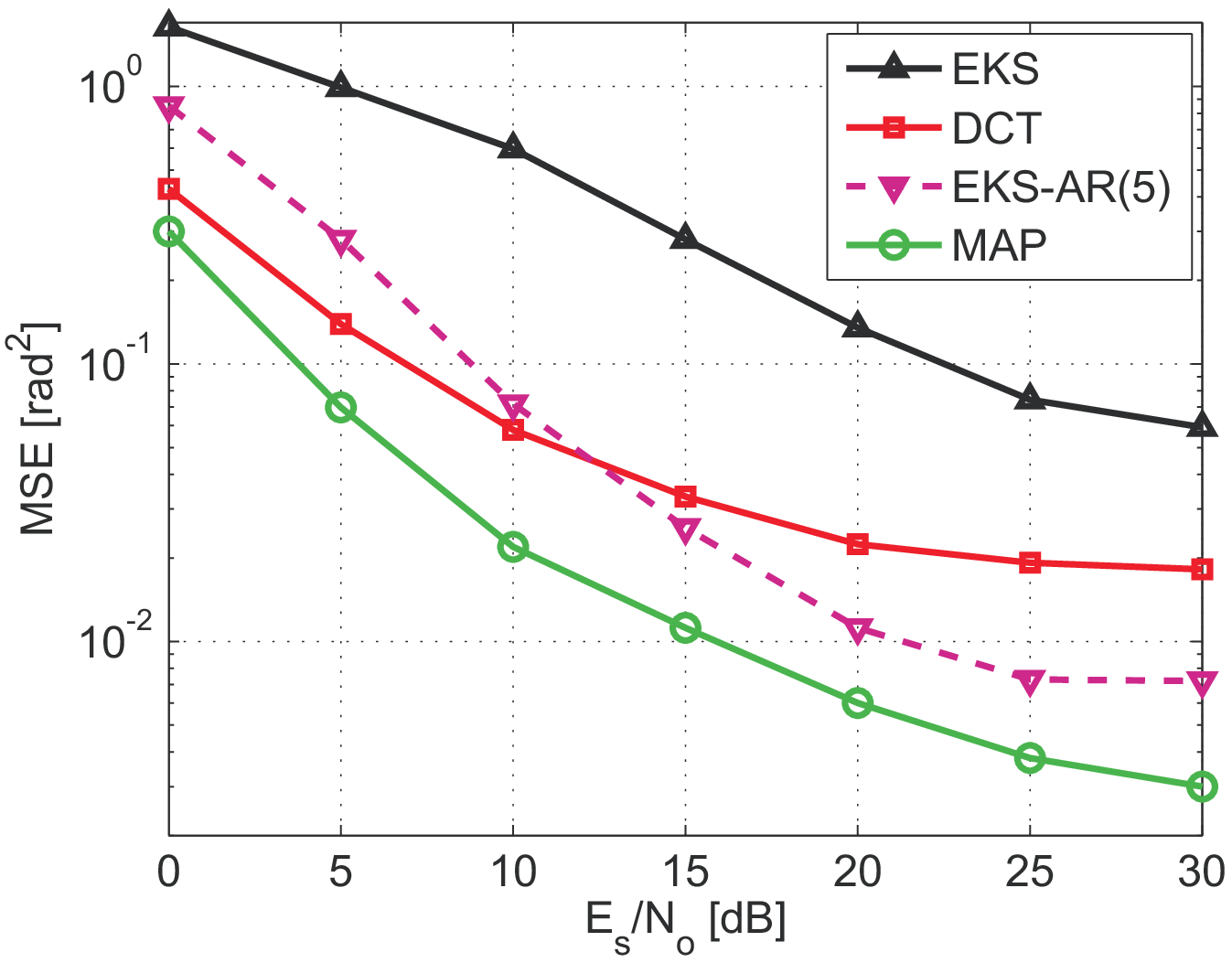}
                \caption{}
                \label{fig:MSE6}
        \end{subfigure}%
        \caption{MSE comparison of different PN estimation methods for different pilot densities over a block of $K=101$ symbols. Phase increment is colored with variance $R_\zeta(0)=10^{-3} \tr{[rad^2]}$. (a) Data-aided case, where all symbols are pilots. (b) Pilot density of $21\%$. (c) Pilot density of $6\%$.}
\label{fig:MSEs}
\end{figure*}
In general, the PN increment process can be modeled with a $p$th-order auto regressive (AR) process as follows
\begin{align}
\label{General_AR_model}
\zeta_{k}=\sum_{i=1}^p\alpha_i\zeta_{k-i}+\Delta_k,
\end{align}
where $\alpha_i$ are the coefficients of the AR model and $\Delta_k$ is modeled as a zero-mean white noise process with variance $\sigma_\Delta^2$. For a given autocorrelation $R_\zeta(l)$ and AR order $p$, the optimal $\alpha_i$ and $\sigma_\Delta^2$ can be computed using algorithms such as the Levinson-Durbin recursion.  We modify the state equation (\ref{Random_Walk}) with the AR model (\ref{General_AR_model}), which results in an augmented state equation,

\begin{align}
\label{augmented_sys_model}
\hspace{-0.3cm}\underbrace{\left[\hspace{-0.2cm}\begin{array}{c}
\theta_k \\ 
\zeta_k \\
\zeta_{k-1} \\
\vdots\\
\zeta_{k-p} \end{array} \hspace{-0.2cm}\right]}_{\triangleq\mathbf{x}_k}=
\underbrace{\left[\hspace{-0.2cm}\begin{array}{c c c c c}
1 & 1 & 0 & \dots & 0\\ 
0 & \alpha_1 & \alpha_2 & \dots & \alpha_p\\
0 & 1 & 0 & \dots & 0\\
\vdots & \ddots & \ddots & \ddots & \vdots\\
0 & \dots & 0 & 1 & 0\end{array} \hspace{-0.2cm}\right]}_{{\triangleq\mathbf{F}}}\underbrace{\left[\hspace{-0.2cm}\begin{array}{c}
\theta_{k-1} \\ 
\zeta_{k-1} \\
\zeta_{k-2} \\
\vdots\\
\zeta_{k-p-1} \end{array} \hspace{-0.2cm}\right]}_{{\triangleq\mathbf{x}_{k-1}}}
+\underbrace{\left[\hspace{-0.2cm}\begin{array}{c}
0 \\ 
\Delta_k \\
0 \\
\vdots\\
0 \end{array}\hspace{-0.2cm} \right]}_{\triangleq\boldsymbol{\Delta_{k-1}}},
\end{align}
where $\mathbf{x}_k$ is the new state vector that has a higher dimension compared to our original state variable $\theta_k$, and $\boldsymbol{\Delta_{k}}$ denotes the new process noise that is white, with covariance
\begin{align}
\label{new_process_noise_covariance}
\mathds{E}[\boldsymbol{\Delta}_{k}\boldsymbol{\Delta}_{k}^T]=\diag\left([0,\sigma^2_\Delta,0,\dots,0]\right).
\end{align}
We can use the new state equation (\ref{augmented_sys_model}) along with the observation model (\ref{System_Model1_Modified}) to estimate $\boldsymbol{\theta}$ by Kalman filtering/smoothing \cite{book_kay_est}. For the colored PN increments with a long memory, normally a high-order AR model is needed that results in a high dimensional state equation (\ref{augmented_sys_model}). In order to reduce complexity, we approximate the colored PN increments with a low-order AR process. Numerical simulations in Sec.~\ref{Sec_Numerical_Results} show that even with such an approximation, the modified extended Kalman smoother (EKS) perform close to the proposed MAP estimator, in several scenarios of interest. 
\vspace{-0.08in}
\section{Numerical Results}
\label{Sec_Numerical_Results}
We now study the performance of our proposed estimators and compare with that of those available in the literature (e.g., \cite{Amblard2003151, Colavolpe2005, Bhatti2009}). We consider transmission of a block of $K=101$ 16-QAM modulated symbols, with uniformly distributed pilots. For simulation of the PN with colored noise increments, we use the results of \cite{Khanzadi2012_1,Khanzadi2013_1}, where the autocorrelation function of the PN increments for oscillators with a colored noise source (flicker noise) has been derived. We set the parameters such that the variance of the PN increments becomes $R_\zeta(l=0)=10^{-3} \tr{[rad^2]}$.

It can be seen in Fig.~\ref{fig:MSEs}-(a) that the MSE of the proposed MAP estimator reaches the BCRB \cite{Khanzadi2013_1} in the data-aided case, where all symbols are pilots. We stop the optimization algorithm when the gradient is sufficiently small (here $|\mathbf{g}(\boldsymbol{\theta})|<10^{-6}$). We observe that the number of required iterations for satisfying any level of accuracy depends on various parameters such as the block length, the PN statistics, the pilot density and the SNR. In general, for most practical scenarios less that $5$ iterations suffice. For instance, for simulations of Fig.~1 with $6\%$ pilot density, at SNRs $0$ and $30$~dB, on average $4.3$ and $2.95$ iterations are required, respectively.  

The MSE of the PN estimator proposed in \cite{Bhatti2009}, based on interpolation of the PN estimates of the pilot symbols using discrete cosine transform (DCT), is close to the MAP estimator in the data-aided case. However, decreasing the pilot density in Fig.~\ref{fig:MSEs}-(b) and (c) to $21\%$ and $6\%$, shows that the DCT-based estimator performs poorly in low-pilot density scenarios. The reason is that this estimator is blind to the statistics of the PN process that limits its performance when the received signals are not reliable.

Using an EKS designed for white PN increments in the colored case results in large MSEs in low-SNR and low-pilot density cases. In high-pilot density scenarios, the observations are reliable and the EKS performs close to the MAP. When the pilot density is low, the tested EKS relies on PN statistics that are not matched to the real process, which results in large MSEs. Now consider the modified EKS designed with the low-order AR approximation. Using a first-order AR model, the modified EKS reaches the MSE of the MAP in the data-aided and $21\%$ pilot density cases. With $6\%$ pilot density, where the modified EKS relies more on the PN statistics, a higher order AR model is needed to improve the performance ($p>5$). Fig.~\ref{fig:MSEs}-(a) also shows the data-aided MSE of the second-order phase tracking loop \cite{Amblard2003151}. 

\begin{figure}[t]
\centering
\includegraphics [width=2.6in]{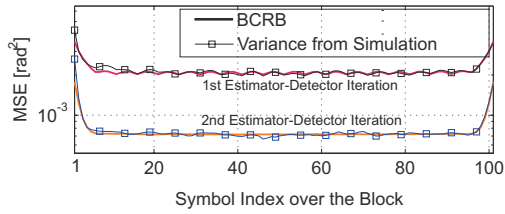}
\caption{Simulated variance of MAP vs. BCRB. Pilot density $21\%$, $R_\zeta(0)=10^{-3} \tr{[rad^2]}$, $\tr{SNR}=20$[dB].}
\label{fig:MAP_var}
\end{figure}

Fig.~\ref{fig:MAP_var} shows that the simulated MSE of the MAP estimator reaches the BCRB. Fig.~\ref{fig:2} and \ref{fig:3} compare the effect of using the discussed estimators on symbol error rate (SER) of the system, after three iterations between the PN estimators and a Euclidian-distance-based symbol detector. Mean and variance of the soft symbols are calculated as the mean and variance of the symbols' a posteriori probabilities. In both scenarios, the MAP estimator outperforms other estimators. It can also be seen that in the $6\%$ density compared to $21\%$ scenario, a higher order AR model  is needed for more accurate approximation of the colored PN increments. In addition to the estimators in Fig.~\ref{fig:MSEs}, we also study the performance of an iterative receiver that is designed based on the sum-product algorithm (SPA) \cite{Colavolpe2005}. This SPA-based receiver performs extremely well in presence of the Wiener PN, but it is not designed for PN with colored increments.
\vspace{-0.3cm}
\section{Conclusions}
In this letter, we showed that deriving the soft-input maximum a posteriori (MAP) estimator for estimation of phase noise (PN) in oscillators with colored noise sources is a concave optimization problem at moderate and high SNRs. Further, we showed that the modified soft-input extended Kalman smoother with low-order AR approximation of the colored PN increments performs close to the MAP in several scenarios. From simulations, we observed that considerable performance gain can be achieved by using the proposed estimators compared to estimators that lack correct statistic of the PN. The gain is more significant in low-SNR or low-pilot density scenarios.
\vspace{-0.5cm}
\begin{appendix}
Here, we find the mean and covariance of the MAP estimation error, defined as $\boldsymbol{\psi}\triangleq(\boldsymbol{\theta^\dagger}-\boldsymbol{\hat{\theta}})$, where $\boldsymbol{\theta^\dagger}$ denotes the true value of $\boldsymbol{\theta}$. We first write the Taylor expansion of $\mathbf{g}(\boldsymbol{\theta})$ around $\boldsymbol{\theta^\dagger}$ and evaluate it at $\boldsymbol{\hat{\theta}}$. Assuming that $\boldsymbol{\hat{\theta}}$ is close to $\boldsymbol{{\theta}}^\dagger$, we can neglect the higher order terms and obtain 
\begin{align} 
\label{Taylor_g_th_at_thh}
\mathbf{g}(\boldsymbol{\hat{\theta}}) \approx \mathbf{g}(\boldsymbol{\theta^\dagger})+\mathbf{H}(\boldsymbol{\theta^\dagger})(\boldsymbol{\hat{\theta}}-\boldsymbol{\theta^\dagger}).
\end{align}
Note that $\boldsymbol{\hat{\theta}}$ is the root of $\mathbf{g}(\boldsymbol{{\theta}})=\mathbf{0}$. Therefore,
\begin{align} 
\label{True_Error_Terms}
\boldsymbol{\theta^\dagger}=\boldsymbol{\hat{\theta}}+
\mathbf{H}^{-1}(\boldsymbol{\theta^\dagger})\mathbf{g}(\boldsymbol{\theta^\dagger}),
\end{align}
where $\boldsymbol{\psi}\triangleq (\boldsymbol{\theta^\dagger}-\boldsymbol{\hat{\theta}})= \mathbf{H}^{-1}(\boldsymbol{\theta^\dagger})\mathbf{g}(\boldsymbol{\theta^\dagger})$ is the estimation error term whose mean is calculated as 
\begin{align}
\label{MAP_Error_Mean}
\mathds{E}[\boldsymbol{\psi}]=\mathds{E}[\mathbf{H}^{-1}(\boldsymbol{\theta^\dagger})\mathbf{g}(\boldsymbol{\theta^\dagger})].
\end{align}
Setting the value of $y_k$ from (\ref{System_Model1_Modified}) in (\ref{d1_log_l}) and (\ref{Hessian_matrix}) gives
\begin{subequations}
\begin{align}
\label{g_def_2}
&\mathbf{g}(\boldsymbol{\theta^\dagger})=2\Big[\Big\{\frac{\Im\{\hat{s}_k^*\check{w}_k\}}{\sigma^2_{k}}\Big\}_{k=1}^K\Big]^T-\mathbf{C}^{-1}\boldsymbol{\theta^\dagger},
\end{align}
\begin{align}
\label{Hessian_matrix_approx_def_2}
&\hspace{-0.35cm}\mathbf{H}(\boldsymbol{\theta^\dagger})=-2\diag\left(\hspace{-0.1cm}\Big[\Big\{\frac{|{\hat{s}}_k|^2+\Re\{ \hat{s}_k^*\check{w}_k\}}{\sigma^2_k}\Big\}_{k=1}^K\Big]\hspace{-0.1cm}\right)-\mathbf{C}^{-1},
\end{align}
\end{subequations}
where $\check{w}_k\triangleq  e^{-j\theta_k} \tilde{w}_k$ with the same statistics as $\tilde{w}_k$. It is clear that $\mathbf{H}(\boldsymbol{\theta^\dagger})$ and $\mathbf{g}(\boldsymbol{\theta^\dagger})$ are independent. Therefore,
\begin{align}
\label{MAP_Error_Mean_final}
\mathds{E}[\boldsymbol{\psi}]=\mathds{E}[\mathbf{H}^{-1}(\boldsymbol{\theta^\dagger})]\mathds{E}[\mathbf{g}(\boldsymbol{\theta^\dagger})]=\mathbf{0},
\end{align}
where the second equality is true because $\mathds{E}[\mathbf{g}(\boldsymbol{\theta^\dagger})]=\mathbf{0}$.

The covariance matrix of $\boldsymbol{\psi}$ is determined as
\begin{align}
\label{MAP_Error_Cov}
\boldsymbol{\Sigma}=\mathds{E}[\boldsymbol{\psi}\boldsymbol{\psi}^T]&=\mathds{E}[\mathbf{H}^{-1}(\boldsymbol{\theta^\dagger})\mathbf{g}(\boldsymbol{\theta^\dagger})\mathbf{g}^T(\boldsymbol{\theta^\dagger})\mathbf{H}^{-1}(\boldsymbol{\theta^\dagger})].
\end{align}
In (\ref{Hessian_matrix_approx_def_2}), it is possible to neglect $\Re\{ \hat{s}_k^*\tilde{w}_k\}$ compared to $|\hat{s}_k|^2$ for moderate and high SNRs. Therefore, $\mathbf{H}(\boldsymbol{\theta^\dagger})$ is approximated as
\begin{align}
\label{Hessian_matrix_approx_1}
\hspace{-0.3cm}\mathbf{H}(\boldsymbol{\theta^\dagger})\approx\mathbf{\tilde{H}}(\boldsymbol{\theta^\dagger})=-2\diag\left(\Big[\Big\{\frac{|\hat{s}_k|^2}{\sigma^2_k}\Big\}_{k=1}^K\Big]\right)-\mathbf{C}^{-1},
\end{align}
which is a deterministic matrix. Thus, the expectation in (\ref{MAP_Error_Cov}) is only over $\mathbf{g}(\boldsymbol{\theta^\dagger})\mathbf{g}^T(\boldsymbol{\theta^\dagger})$. 
Based on (\ref{g_def_2}) and after straightforward mathematical manipulation, 
\begin{align}
\label{Exp_ggt}
\mathds{E}[\mathbf{g}(\boldsymbol{\theta^\dagger})\mathbf{g}^T(\boldsymbol{\theta^\dagger})]=-\mathbf{\tilde{H}}(\boldsymbol{\theta^\dagger}),
\end{align}
and hence, $\boldsymbol{\Sigma}\approx-\mathbf{\tilde{H}}^{-1}(\boldsymbol{\theta^\dagger})$. Employing the data-aided BCRB for estimation of colored PN derived in \cite{Khanzadi2013_1}, and using the system model (3), it is straightforward to find the soft-input BCRB, and show that it is identical to the covariance of estimation error $\mathbf{\Sigma}$.
\begin{figure}[t]
\centering
\includegraphics [width=2.5in]{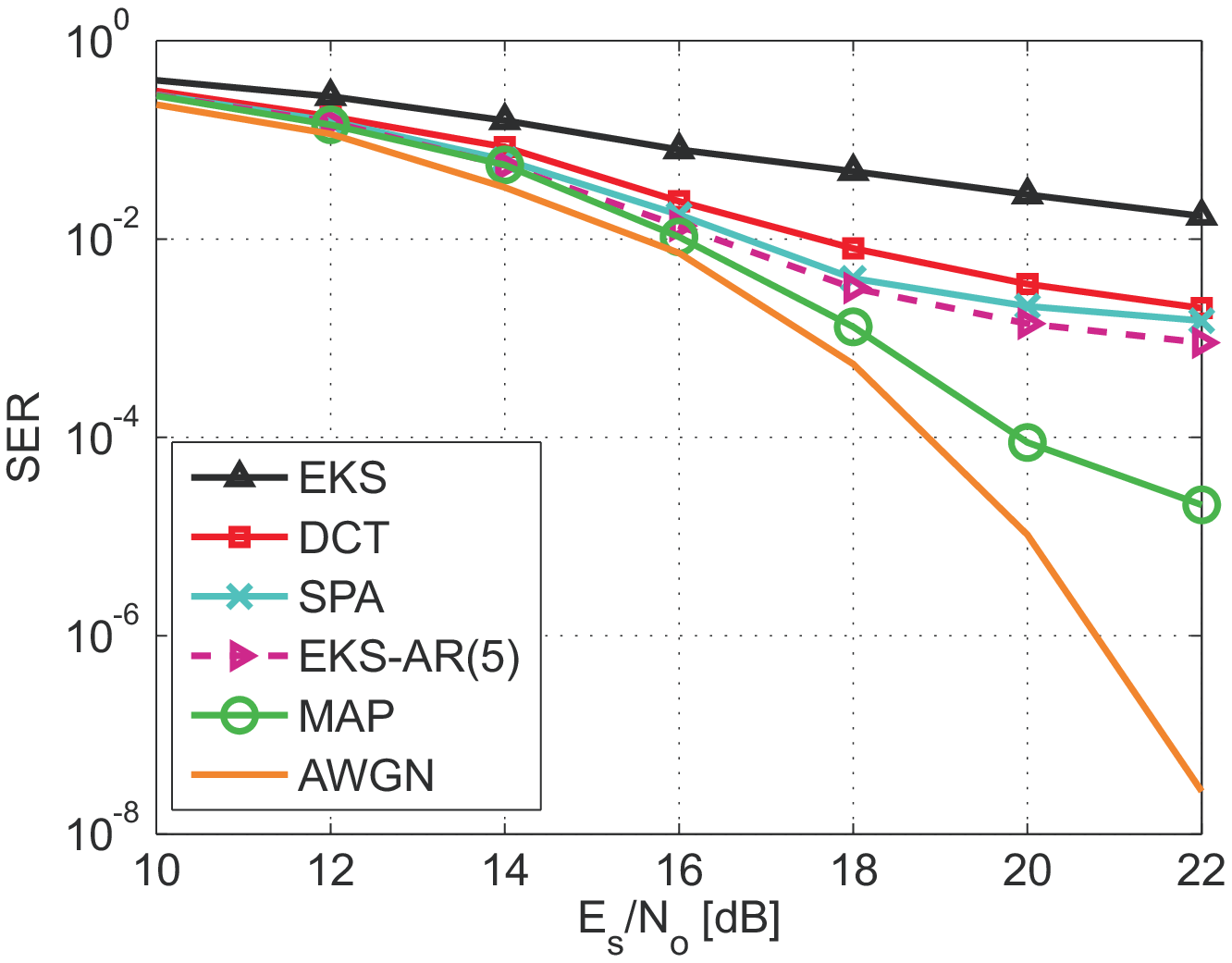}
\caption{SER comparison of different PN estimation methods after three estimation-detection iterations. Pilot density $6\%$ and $R_\zeta(0)=10^{-3} \tr{[rad^2]}$.}
\label{fig:2}
\end{figure}
\begin{figure}[t]
\centering
\includegraphics [width=2.5in]{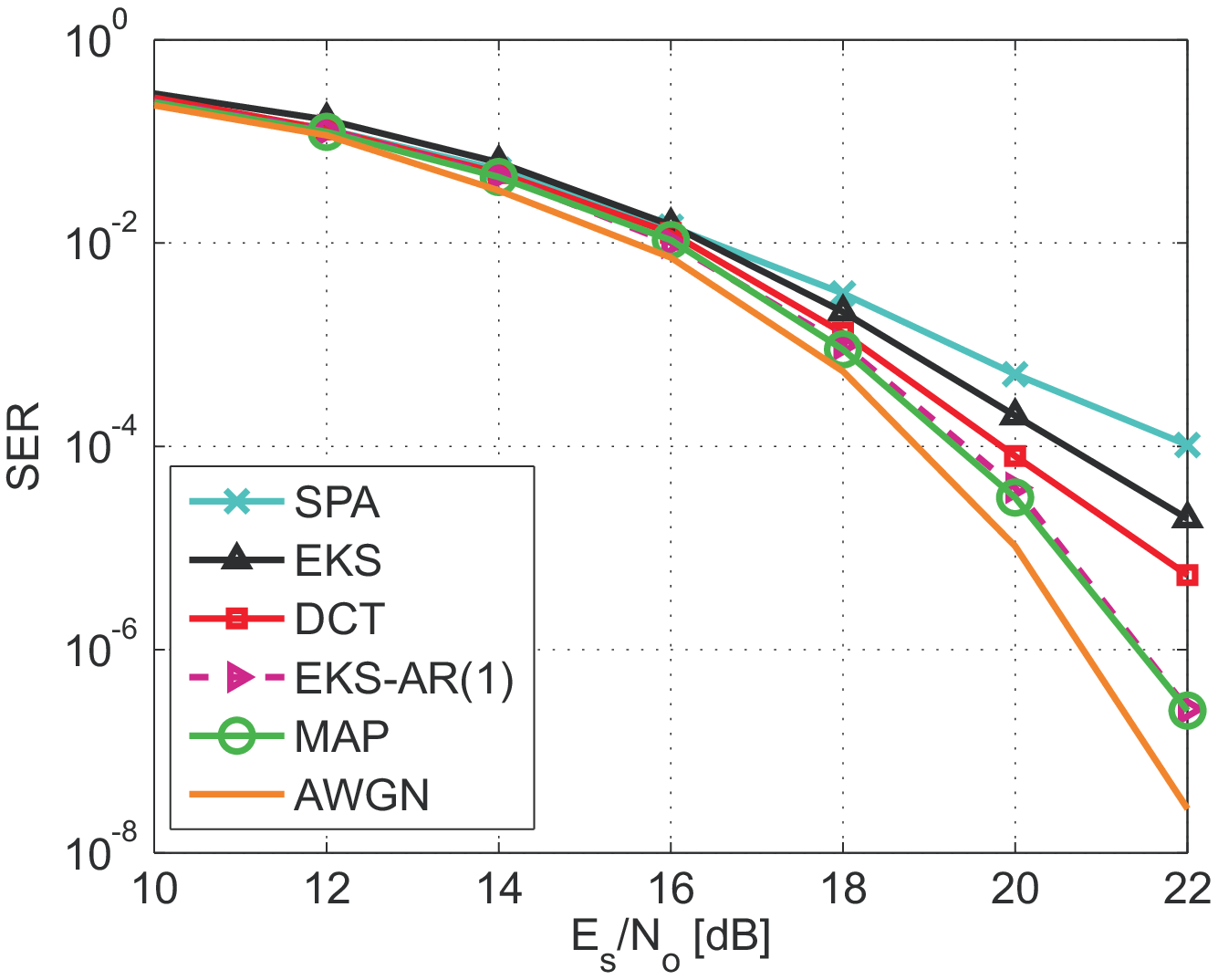}
\caption{SER comparison of different PN estimation methods after three estimation-detection iterations. Pilot density $21\%$ and $R_\zeta(0)=10^{-3} \tr{[rad^2]}$.}
\label{fig:3}
\end{figure}
\end{appendix}
\vspace{-0.4cm}
\bibliographystyle{IEEEtran}
\bibliography{IEEEabrv,references}
\end{document}